\documentclass[sigconf,nonacm]{acmart}
\usepackage{stfloats} 
\usepackage[dvipsnames,table]{xcolor}

\usepackage[utf8]{inputenc}  
\usepackage{alltt}
\usepackage{wrapfig}
\usepackage{multicol}
\usepackage{enumitem}
\setlist[itemize]{leftmargin=1em}
\usepackage{array}
\newcolumntype{C}[1]{>{\centering\arraybackslash}p{#1}}
\usepackage[ruled,vlined]{algorithm2e}

\setcitestyle{square}

\newcommand{\tao}[1]{\textcolor{teal}{#1}}

 \newcommand{\rev}[1]{#1}

\usepackage{etoolbox}

\AtBeginDocument{}

\copyrightyear{2026}
\acmYear{2026}
\setcopyright{cc}
\setcctype{by}
\acmConference[MSR '26]{23rd International Conference on Mining Software Repositories}{April 13--14, 2026}{Rio de Janeiro, Brazil}
\acmBooktitle{23rd International Conference on Mining Software Repositories (MSR '26), April 13--14, 2026, Rio de Janeiro, Brazil}
\acmPrice{}
\acmDOI{10.1145/3793302.3793306} 
\acmISBN{979-8-4007-2474-9/2026/04}
\makeatletter
\patchcmd{\maketitle}{\@copyrightspace}{}{}{}

\makeatother
\begin{document}
\title{MOOT: a Repository of Many Multi-Objective Optimization Tasks}
 \author{Tim Menzies}
\affiliation{\institution{Computer Science\\ NC State,  USA \country{}\\ {{\tt timm@ieee.org}}}}

\author{Tao Chen \\Yulong Ye }
\affiliation{\institution{Computer Science \\University of Birmingham, UK \country{} {{\tt t.chen@bham.ac.uk yxy382@student.bham.ac.uk }}}}

\author{Kishan Kumar Ganguly, Amirali Rayegan, Srinath Srinivasan, Andre Lustosa}
\affiliation{\institution{Computer Science, NC State, USA \country{}\\{{\tt \{kgangul; arayega; ssrini27; alustos\}@ncsu.edu;  }}}}

\authornote{All authors contributed equally to this research.}

\renewcommand{\shortauthors}{Menzies, Chen et al.}
 
\begin{abstract}
Software engineers must make decisions that  trade off competing goals (faster vs. cheaper, secure
vs. usable, accurate vs. interpretable, etc.). Despite MSR's proven
techniques for exploring such goals, researchers still
struggle with these trade-offs. Similarly, industrial practitioners 
 deliver
sub-optimal products since they lack the tools needed to explore
these trade-offs.

 To address this, MOOT (http://tiny.cc/moot) is
a repository of many  SE multi-objective optimization tasks.
MOOT's 120+ tasks cover software configuration, cloud tuning, project health, process modeling, hyperparameter optimization, and more. 

Sample scripts for reading MOOT and generating baseline results are available-- just clone the repository and run the  sample rqx.sh files (from
tiny.cc/moot0). 

To the best of our knowledge, MOOT is the largest and most
varied collection of real multi-objective optimization tasks in SE.
\rev{We note that MOOT's novelty is infrastructural, not algorithmic---we contribute curated data and research enablement, not new optimization methods.}
MOOT enables harder and more credible research. MOOT lets us
replace studies on  toy problems 
(or  just half a dozen hand-picked examples) with
case studies on  120+  examples. Such studies could focus on stability, sample efficiency, failure modes,  cross-domain generality,
or many other questions (see list in this document). 
 
\end{abstract}

\received{23 Oct 2025}

\maketitle


\section{Introduction}\label{why}



\begin{table*} 
\scriptsize
\renewcommand{\baselinestretch}{0.9}
\caption{Tasks in MOOT. ``x/y'' denotes the number of independent and
dependent attributes, For an example of how each dataset looks like refer to Figure~\ref{moot}.   
These datasets come from papers published in top SE venues such as 
the International Conference on Software Engineering \cite{chen2026promisetune, DBLP:conf/icse/WeberKSAS23,10172849,DBLP:conf/icse/HaZ19},
Foundations of SE (FSE) conference~\cite{nair2017using,DBLP:conf/sigsoft/JamshidiVKS18}
IEEE Trans. SE \cite{chen2025accuracy,xia2020sequential,krishna2020whence,Chen19,krall2015gale},
the Information Software Technology journal \cite{chen2018beyond,fu2016tuning},
Empirical Softw. Eng. \cite{hulse2025shaky, peng2023veer,guo2018data},
Mining Software Repositories \cite{nairMSR18},
IEEE Access \cite{lustossa2024isneak},
 ACM Trans. SE Methodologies~\cite{lustosa2024learning} and the Automated Software Engineering Journal~\cite{nair2018faster}.}
\label{datasets-summary}
\begin{tabular}{@{}C{1.1cm}@{~}p{2.3cm}p{3.6cm}p{4.2cm}p{1cm}p{1.2cm}p{2.6cm}@{}}
\# Datasets & Dataset Type                     & File Names                                                               & Primary Objective                                                     & x/y          & \# Rows & Cited By       \\ \midrule
25          & \begin{tabular}[c]{@{}l@{}}Specific Software\\Configurations\end{tabular} & SS-A to SS-X, billing10k                                                 & Optimize software system settings                                     & 3-88/2-3   & 197–86,059 & \cite{Amiraliminimaldata, menzies2025the, Kishanbingo, lustossa2024isneak, senthilkumar2024can, lusstosa2025less, nairMSR18, peng2023veer,nair2017using,nair2018faster}   \\ 
12          & \begin{tabular}[c]{@{}l@{}}PromiseTune Software\\Configurations\end{tabular} & \begin{tabular}[c]{@{}l@{}} 7z, BDBC, HSQLDB, LLVM, PostgreSQL, \\ dconvert, deeparch, exastencils, javagc, \\ redis, storm, x264\end{tabular}                                                 & Software performance optimization                                     &  9-35/1  &  864-166,975 &  \cite{chen2026promisetune, chen2025accuracy, DBLP:conf/icse/XiangChen26, DBLP:journals/pacmse/Gong024, gong2024dividable} \\ \midrule
1           & Cloud                            & HSMGP num                                                                & Hazardous Software, Management Program data                           & 14/1         & 3,457 &  \cite{Amiraliminimaldata, menzies2025the, chen2025accuracy, Kishanbingo, senthilkumar2024can}     \\
1           & Cloud                            & Apache AllMeasurements                                                   & Apache server performance optimization                                & 9/1          & 192  &    \cite{Amiraliminimaldata, menzies2025the, chen2025accuracy, Kishanbingo, senthilkumar2024can}     \\
1           & Cloud                            & SQL AllMeasurements                                                      & SQL database tuning                                                   & 39/1         & 4,654 &    \cite{Amiraliminimaldata, menzies2025the, Kishanbingo, senthilkumar2024can}    \\
1           & Cloud                            & X264 AllMeasurements                                                     & Video encoding optimization                                           & 16/1         & 1,153  &   \cite{Amiraliminimaldata, menzies2025the, Kishanbingo, senthilkumar2024can}   \\
7           & Cloud                            & (rs—sol—wc)*                                                             & misc configuration tasks                                              & 3-6/1      & 196–3,840 &  \cite{Amiraliminimaldata, menzies2025the, Kishanbingo, senthilkumar2024can, lusstosa2025less, nairMSR18}  \\ \midrule
35          & Software Project Health          & Health-ClosedIssues, -PRs, -Commits                                      & Predict project health and developer activity                         & 5/2-3      & 10,001    &   \cite{Amiraliminimaldata, menzies2025the, Kishanbingo, senthilkumar2024can, lusstosa2025less,lustosa2024learning} \\ \midrule
3           & Scrum                            & Scrum1k, Scrum10k, Scrum100k                                             & Configurations of the scrum feature model                             & 124/3      & 1,001–100,001 & \cite{Amiraliminimaldata, menzies2025the, lusstosa2025less, lustossa2024isneak} \\ \midrule
8           & Feature Models                   & FFM-*, FM-*                                                              & Optimize number of variables, constraints and Clause/Constraint ratio & 128-1,044/3 & 10,001  &  \cite{Amiraliminimaldata, menzies2025the, lusstosa2025less, lustossa2024isneak}    \\ \midrule
1 &	Software Process Model &	nasa93dem &	Optimize effort, defects, time and LOC	& 24/3 &	93  & \cite{menzies2025the, senthilkumar2024can, lusstosa2025less, lustosa2024learning}\\
1           & Software Process Model           & COC1000                                                                  & Optimize risk, effort, analyst experience, etc                        & 20/5         & 1,001    &  \cite{Amiraliminimaldata, menzies2025the, senthilkumar2024can, lustosa2024learning,chen2018beyond}   \\
4           & Software Process Model           & POM3 (A–D)                                                               & Balancing idle rates, completion rates and cost                       & 9/3          & 501–20,001   & \cite{menzies2025the, senthilkumar2024can, lusstosa2025less, lustosa2024learning, lustossa2024isneak}\\
4           & Software Process Model    & XOMO (Flight, Ground, OSP)                                               & Optimizing risk, effort, defects, and time                            & 27/4         & 10,001    &  \cite{menzies2025the, senthilkumar2024can, lusstosa2025less, lustosa2024learning, lustossa2024isneak,chen2018beyond}  \\ \midrule
3           & Miscellaneous                             & auto93, Car\_price, Wine\_quality                                        & Miscellaneous                                                         & 5-38/2-5   & 205–1,600  &  \cite{Amiraliminimaldata, menzies2025the, Kishanbingo, senthilkumar2024can, lusstosa2025less, lustosa2024learning} \\ \midrule
4           & Behavioral                       & all\_players, student\_dropout,\newline HR-employeeAttrition, player\_statistics & Analyze and predict behavioral patterns                              & 26-55/1-3  & 82–17,738   &   From  \cite{nyagami_fc25_kaggle_2025, abdullah0a_student_dropout_analysis_prediction_2025, die9origephit_fifa_wc_2022_complete_2025, pavansubhasht_ibm_hr_analytics_attrition_2025}\\ \midrule
4           & Financial                        & BankChurners, home\_data, Loan, \newline Telco-Churn                              & Financial analysis and prediction                                     & 19-77/2-5  & 1,460–20,000 &  From \cite{blastchar_telco_customer_churn_2025, lorenzozoppelletto_financial_risk_for_loan_approval_2025, dansbecker_home_data_for_ml_course_2025, sakshigoyal7_credit_card_customers_2020} \\ \midrule
3           & Human Health Data                & COVID19, Life\_Expectancy, \newline hospital\_Readmissions                        & Health-related analysis and prediction                                & 20-64/1-3  & 2,938–25,000 &   From \cite{dansbecker_hospital_readmissions_2025, kumarajarshi_life_expectancy_who_2025, hendratno_2022}  \\ \midrule
2           & Reinforcement Learning           & A2C\_Acrobot, A2C\_CartPole                                              & Reinforcement learning tasks                                          & 9-11/3-4   & 224–318  &     \\ \midrule
5           & Sales                            & accessories, dress-up, Marketing\_Analytics, socks, wallpaper            & Sales analysis and prediction                                         & 14-31/1-8  & 247–2,206 &   From \cite{jessicali9530_animal_crossing_new_horizons_nookplaza_dataset_2021, jackdaoud_marketing_data_2022, syedfaizanalii_car_price_dataset_cleaned_2025}  \\ \midrule
2	& Software testing	& test120, test600	& Optimize the class	& 9/1	& 5,161\\ \midrule

127         & \textbf{Total}                            &                                                                          &                                                                       &              &        &      
\end{tabular}
\end{table*}

The MOOT repository~\cite{mootrepo}\footnote{URL: \url{http://github.com/timm/moot};\newline
DOI: \url{https://doi.org/10.5281/zenodo.17354083};\newline
ZENODO: \url{https://zenodo.org/records/17354083}}
enables data mining research for trading off multiple user goals  such as {\em better, faster, cheaper} (fewer bugs, less time, less cost); {\em performance vs. sustainability} (faster response, less energy); and {\em hyperparameter optimization} (finding learner parameters that minimize false alarms and maximize recall). Formally, these are multi-objective {\em configuration optimization tasks}.

MSR has   recognized configuration optimization as promising for data mining. A 2019 NII Shonan Meeting on Data-driven Search-based SE\footnote{\url{http://goo.gl/f8D3EC}} drew two dozen+ senior MSR researchers who concluded that repository mining can benefit from MSR techniques (e.g. tuning MSR algorithm parameters~\cite{fu2016tuning}). Conversely, MSR's data miners help optimization (e.g. summarize data to speed optimization~\cite{krall2015gale}).

Yet this convergence has not gained momentum. In 2025, researchers still use simplistic hyperparameter methods~\cite{ali2024enhancing} (e.g. grid search with one loop per hyperparameter\footnote{Grid search is adequate for tuning a few parameters, but much better alternatives exist~\cite{bergstra2011algorithms} with public implementations~\cite{bergstra2013hyperopt,akiba2019optuna,awad2021dehb}.}). Meanwhile, configuration optimization has suffered from external validity since much prior work was tested on very little data\footnote{As reported in Table~2 of \url{https://arxiv.org/pdf/2501.00125}, much prior work has tested their optimization methods on five data sets or less.}. All this amounts to a critical missed opportunity for the MSR community.

\begin{figure}[!t] 
\begin{center} \includegraphics[width=\linewidth]{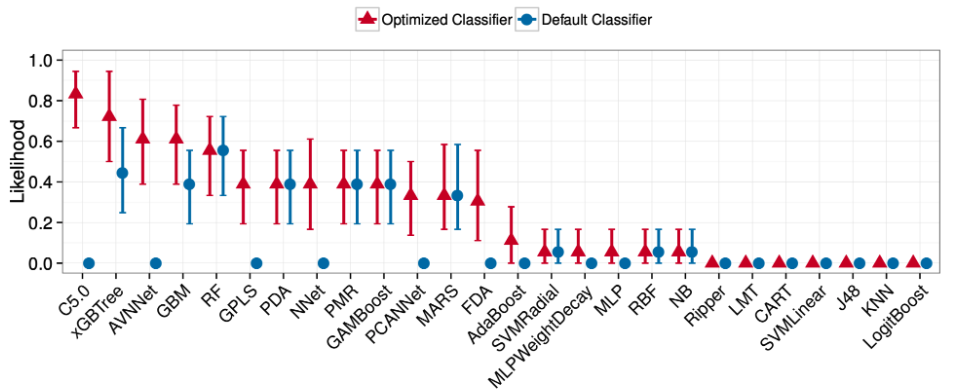}\end{center}
\caption{Odds of defect predictor performing best. \textcolor{blue}{BLUE} = pre-optimization, \textcolor{red}{RED} = post-optimization. From~\cite{Tantithamthavorn16}.}
\label{tant}
\end{figure}
 
Why should MSR use MOOT?  Firstly,
\textbf{automatic configuration dramatically improves lear\-ner performance.} A staple of MSR research is software analytics, where data miners are applied to software project data. For software analytics, automatic configuration optimization of learner parameters can dramatically transform results~\cite{fu2016tuning, agrawal2019dodge, yedida2021value, yedida2023find}. Figure~\ref{tant} shows an example of defect prediction optimization (\textcolor{blue}{BLUE}=~pre-optimization, \textcolor{red}{RED}=~post-optimization). When optimized, the C5.0 algorithm (left side) moved from being the {\em worst} to being the {\em best} learner. Similar gains have been seen in other SE domains~\cite{agrawal2018better,agrawal2019dodge,Tantithamthavorn16,fu2016tuning}. These improvements mean unoptimized data miner results can be easily refuted by tuning\footnote{E.g. ~\cite{agrawal2018wrong} showed (a)~half of prior results were just learner nondeterminism; (b)~optimization can tame that nondeterminism.}. MSR needs to understand how much optimization changes conclusions from off-the-shelf tools.

Secondly, \textbf{industry urgently needs better configuration methods.} Modern software has extensive configurability, but tuning parameters critically affect performance~\cite{DBLP:journals/tosem/ChenL23a}. For example, \textsc{Storm}'s defaults yield $480\times$ performance degradation vs. optimal parameters~\cite{DBLP:conf/mascots/JamshidiC16}. Such poor performance is hardly surprising since industrial optimization can
result in  suboptimal products~\cite{Krishna:2016}:
\begin{itemize}
    \item Configuration spaces explode exponentially (in \textsc{7z}: 14 parameters = one million configurations).
    \item Performance landscapes are rugged and sparse~\cite{DBLP:journals/tosem/GongC25,lustosa2024learning,chen2026promisetune}, creating local optima traps.
    \item Evaluation is costly: \textsc{x264}'s 11 parameters need $1,536$ hours to explore~\cite{DBLP:conf/wosp/ValovPGFC17}, limiting budgets to dozens of evaluations~\cite{DBLP:journals/tse/Nair0MSA20,DBLP:conf/icse/0003XC021}.
\end{itemize}
Thirdly, \textbf{MSR algorithms excel on large data spaces~\cite{guocart}.} Configuration optimization seeks $c^* \in \mathcal{C}$ optimizing multiple objectives (e.g., {\em maximize} database throughput, {\em minimize} energy). Given $f: \mathcal{C} \rightarrow \mathbb{R}^M$ mapping configurations to performance metrics:
\[c^* = \operatorname{argmax}_{c \in \mathcal{C}} f(c)\]
The space of possible configurations seem too large to explore. For example, MySQL's 460 binary options generate $2^{460}$ configurations---more than the $2^{80}$ stars in the observable universe~\cite{doe2023personal}. But in the past, MSR researchers have successfully tackled analogous large-space problems. Yu et al. showed vulnerability detection across tens of thousands of reports needed only hundreds of support vectors~\cite{yu2019improving}, which is to say that buried inside seemingly complex configuration spaces contain exploitable and simpler structure.

\textbf{Enter MOOT.} To address this critical opportunity, we built MOOT, a curated repository of configuration and optimization data drawn from top SE venues (e.g. ICSE, FSE, TSE, IST, EMSE, TOSEM, ASE, etc. \cite{chen2026promisetune, DBLP:conf/icse/WeberKSAS23,10172849,DBLP:conf/icse/HaZ19,nair2017using,DBLP:conf/sigsoft/JamshidiVKS18,chen2025accuracy,xia2020sequential,krishna2020whence,Chen19, krall2015gale,chen2018beyond,fu2016tuning,hulse2025shaky, peng2023veer,guo2018data,lustosa2024learning,nair2018faster}). Using this resource, researchers have already produced new state-of-the-art techniques for SE optimization, software configuration, cloud tuning (Apache, SQL, X264), project health prediction, feature models, process modeling (nasa93dem, COC1000, POM3, XOMO), behavioral analytics, financial risk, churn prediction, health data, reinforcement learning, sales forecasting, testing, text mining, and more. With MOOT, these ideas (previously validated on a few datasets) can now be evaluated at scale, opening the door to more research questions (see end of this article), broader validation, replication, and discovery of more general principles. For an example of this process, see \S\ref{eg} where clustering methods enable very fast optimization.
 
{\bf In summary:} researchers and industry have complex configuration problems which MSR can solve with data mining.  MOOT is one way to explore such problems.

\section{Inside MOOT}\label{what}
Table~\ref{datasets-summary} showed the current 120+   tasks in MOOT.  
\rev{When reading that table, note that datasets are grouped by domain; the ``Cited By'' column shows prior work using each dataset.}
Each task has:
\begin{itemize}
\item
One to 11 goals (median=3)
\item
3 to 10044 input variables (median=11) 
\item and   100 to over 100,000 instances (median=10,000).  
\end{itemize}
 The last few rows of Table~\ref{datasets-summary} show  non-SE datasets. These  are useful
 for explaining  MOOT to visitors from other fields of research. 
\rev{All datasets are reformatted to a common schema (described below), versioned on GitHub, and archived on Zenodo.
While many datasets originate from work by the authors, all come from peer-reviewed publications at top SE venues.}

To the best of our knowledge, MOOT is the largest and most varied
collection of real multi-objective optimization tasks in SE. Earlier
resources (e.g., SPLOT) were valuable but narrow and are now offline.
Toolkits like Pygmo or Platypus offer synthetic benchmarks like ZDT,
DTLZ, Rosenbrock's banana\footnote{
The banana is a task with a non-convex decision frontier; i.e.
minimize $\sum_{i=1}^{n-1} \left[ 100(x_{i+1} - x_i^2)^2 + (x_i - 1)^2
\right], \quad -2.048 \leq x_i \leq 2.048, \quad i = 1, \ldots, n$.
}. But it's somewhat bananas to expect synthetic problems to convince
hard-nosed business users. MOOT's datasets come from published SE
studies, real performance logs, cloud systems, defect predictors, and
tuning tasks where bad configurations cost time, money, and
credibility.

Fig.~\ref{moot}
shows the  typical structure  of a MOOT dataset. 
In Fig.~\ref{moot}, the goal  is to tune {\em Spout\_wait, Spliters, Counters} in order to achieve the best {\em
Throughput/Latency}.  
As seen in that figure, 
\begin{enumerate}
    \item MOOT datasets are tables with $x$ inputs and $y$ goals. 
    \item The first row shows the column names.
    \item The other rows show   examples of a $y=f(x)$ relation.
    \item Numerics   start with   uppercase letters, all else are symbolic.
    \item Goal columns (e.g. Fig.~\ref{moot}'s {\em Throughput+, Latency-}) use +/- to denote maximize and minimize.
    \item  Columns with uppercase "X" at the end of their names are to be ignored by the inference.
\end{enumerate}
For a larger example of MOOT data, see the repository itself; e.g \url{https://github.com/timm/moot/blob/master/optimize/config/SS-M.csv}.
For the purposes of illustration, the rows in Figure~\ref{moot}  are sorted
from best to worst based on those goals. During experimentation, row order should initially be randomized.

\begin{figure}[!t]
\caption{An example of a MOOT dataset.}\label{moot}
{\scriptsize
 
 \begin{alltt}
            x = independent values          | y = dependent values
            --------------------------------|----------------------
            Spout_wait, Spliters, Counters, | Throughput+, Latency-
               10,        6,       17,      |    23075,    158.68
                8,        6,       17,      |    22887,    172.74 
              [Skipped],  ...,      ...,           ...,    ... 
            10000,        1,       18,      |   402.53,    8797.5 
            10000,        1,        1,      |   310.06,    9421
\end{alltt}
}
 \end{figure}

\section{  Using MOOT: an Example}\label{eg}
 
 MOOT supports many research questions (see Table~\ref{fig:research-roadmap}).
 This section offers baseline
 results of one of those questions.
 
Consider the problem of making conclusions via labeled data.
{\bf Labeling by human experts}  is possible but slow and error-prone when rushed~\cite{easterby1980design}, often taking hours for just a few cases~\cite{KingtonAlison2009,lustosa2024learning,valerdi2010heuristics}.
{\bf Historical logs} provide large label sets but can be unreliable (e.g., 90\% of technical debt “false positives” were incorrect~\cite{yu2020identifying}; similar issues occur in security~\cite{wu2021data}, static analysis~\cite{kang2022detecting}, and defect data~\cite{shepperd2013data}).
{\bf Automated labeling} also faces limits: regex-based heuristics are crude~\cite{kamei2012large}; LLMs are only assistive (but not authoritative).
Even in domains with naturally occurring oracles (e.g. compile with certain Makefile settings; run the full test suite) but, as noted, these can be extremely slow (e.g. recall the x264 example~\cite{DBLP:conf/wosp/ValovPGFC17}). 

 When we cannot trust many labels, we must do what we can with very few labels. Enter data mining.
 The {\tt rq2.sh} script of  \url{http://tiny.cc/moot0}  clusters data, labels a few items per cluster, then uses that information to guess the labels of nearby items. 
 We make no claim that this {\bf BASELINE}  MOOT optimizer is state-of-the-art. 
 We only show it here as an example of   optimize-via-data-mining (in this case, Euclidean distance measures and clustering).
 \rev{This baseline is deliberately naive, chosen for pedagogical simplicity rather than performance; it represents common practice, not a competitive benchmark.}

After  dividing data 50:50 into {\em train} and {\em hold-out}, 
 {\bf BASELINE}:
\begin{itemize}
\item {\bf TRAINS   a model:}
  \begin{itemize}
  \item Label, say, $n_1=30$ randomly selected training examples;
  \item Sort them on $y$.
  \item Divide   sorted   data into a $\sqrt{n}$ best set and a $n - \sqrt{n}$ rest set
  \item Return $\mathit{Model}(\mathit{row})=\mathit{dist}(\mathit{row},C_b) - \mathit{dist}(\mathit{row},C_r)$ where $C_b$ and $C_r$ are the centroids on best and rest sets; and $\mathit{dist}$ is Euclidean distance between $x$ values
  \end{itemize}

\item {\bf TESTS a model on hold-outs:}
  \begin{itemize}
  \item Sort the hold-out on $y$ using $\mathit{Model}$
  \item Label the first, say, $n_2=10$ items in that sort
  \item Return the best of these as the recommendation
  \end{itemize}
\end{itemize}
 Note that this process uses only $(n_1=30)+(n_2=10)$ labels. 
\rev{The train/test split, distance metric, and labeling budget ($n_1$, $n_2$) are design choices, not constraints imposed by MOOT.}

 \definecolor{myred}{HTML}{E93034} 
 \begin{figure}[!b]
\begin{minipage}{1.8in}
\includegraphics[width=1.5in]{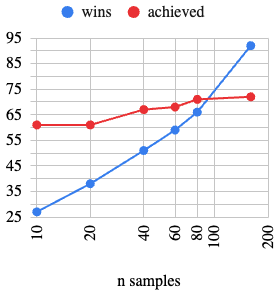}
\end{minipage} \begin{minipage}{1.4in}
\footnotesize 
For the  \textcolor{blue}{{\bf BLUE}} curve,
 $y$ is the
 percent frequency where  
$x < 160$ samples performed statistically as well as 160 samples.\\~\\
For the \textcolor{myred}{{\bf RED}} curve,  $y$  shows
the performance of the row
recommended by {\bf BASELINE}
after $x$ samples.
This number is normalized
such that  $y=0$ means  ``same
as the untreated mean goal score''
(i.e. no improvement); and $y=100$
means ``best possible row found''.
\end{minipage}
\caption{Mean results  (20 repeats), {\bf BASELINE}   on 127  datasets.}\label{results}
 \end{figure}

Figure~\ref{results} compares using only $n$ labels versus all $N>n$, evaluated with nonparametric tests (Kolmogorov--Smirnov for significance and Cliff’s Delta for effect size, see MOOT0’s \texttt{stats.py}).

The \textcolor{blue}{\textbf{BLUE}} curve shows  the usual story; i.e. more labels ($N>n$)  wins statistically. But the \textcolor{myred}{\textbf{RED}} curve shows something
unexpected: after about 40 labels, there  is very little further improvement (from 68 to 73). For mission-critical and safety-critical applications, that improvement may be important. But for many other applications, Figure~\ref{results} 
suggests that a mere 40 labels could suffice.

Note that this result runs contrary to much of the current fashion in AI  since it says   more data is not always better. If this holds for other optimizers (and not just this simplistic   \textbf{BASELINE} method)
then this would be  an important insight   for resource-limited processes such e.g., cutting cloud costs, deploying on the edge, or avoiding fatigue in human-in-the-loop labeling.

\begin{table*}[!t]
\footnotesize
\caption{\rev{Research questions: a roadmap for future work using MOOT data. 
Sections marked \textsc{[n]} are near-term (immediately actionable with current MOOT data); 
sections marked \textsc{[e]} are exploratory (longer-term or speculative directions).}}
\label{fig:research-roadmap}
\begin{minipage}[t]{0.48\textwidth}
\begin{tabular}{@{} p{\columnwidth} @{}}
\hline  
{\bf A.   Optimization Strategies \& Performance} \rev{\textsc{[n]}}: 
Best way to find good solutions? \\
  
\begin{itemize}[leftmargin=*, noitemsep, topsep=1pt, parsep=0pt]
    \item Sample size
    \begin{itemize}[leftmargin=*, noitemsep, topsep=1pt, parsep=0pt]
        \item \textbf{Minimality:} What is the minimum 
        budget for reliable optimization?
        \item \textbf{Surprising Simplicity:} Preliminary results (e.g. 
        Figure~\ref{results}) suggest optimization in SE is 
        surprisingly simple. What makes SE problems unique and so 
        simple?
    \end{itemize}
    \item Optimization:
    \begin{itemize}[leftmargin=*, noitemsep, topsep=1pt, parsep=0pt]
        \item \textbf{Algorithms:} Which AI optimization 
        algorithms are best for MOOT data?
        \item \textbf{Baselines:} If publishing on MOOT, what are 
        the state of the art algorithms against which we 
        should compare?
        \item \textbf{Strategy:} When is it best to use 
        {\em "Pool-based"} search of pre-enumerated examples or 
        {\em "Query-based"} methods that interpolate between 
        existing examples?
        \item \textbf{Landscape:} Are optimization results controlled by 
        algorithms? Or if controlled by the overall ``shape'' 
        of the data, do we need different ``landscape 
        aware'' optimizers?
    \end{itemize}
    \item Metrics:
    \begin{itemize}[leftmargin=*, noitemsep, topsep=1pt, parsep=0pt]
        \item \textbf{x-metrics:} How do distance metrics 
        (Euclidean, Hamming) affect performance, and can we 
        predict the best?
        \item {\bf y-metrics:} What measures (e.g., 
        "distance to heaven", Chebyshev, HV, IGD) best identify good 
        solutions?
        \item \textbf{Surrogates:} If evaluating via quickly built 
        approximations (e.g. random forests), how to certify 
        the surrogates? How good do they need to be?
    \end{itemize}
    \item \textbf{Ensembles:} Can ensemble of algorithms or ensembles 
    built over bagged data outperform single-algorithm approaches?
    \item \textbf{Causality:} Does using causality
    (rather than just correlation) improve optimization?
\end{itemize}
\vspace{-0.2cm}
\noindent\rule{\linewidth}{0.4pt}
~\\

{\bf B. Human Factors \& Interpretability} \rev{\textsc{[n]}}: 
Are solutions understandable, usable? \\
 
\addlinespace[1pt]
\begin{itemize}[leftmargin=*, noitemsep, topsep=1pt, parsep=0pt]
    \item \textbf{Explanation:} How can we (visually or otherwise) 
    explain opaque optimization results to help stakeholder 
    decisions?
    \item \textbf{Trade-offs:} What analytics best help 
    humans understand the decision space (especially for 5+ 
    objectives)?
    \item \textbf{Requirements engineering:}
    Using MOOT, can we test requirements engineering methods 
    to help stakeholders trade off between their different 
    concerns?
    \item Bias:
    \begin{itemize}[leftmargin=*, noitemsep, topsep=1pt, parsep=0pt]
        \item \textbf{Unfairness:} Optimization is always biased 
        towards the stated goals. Does that routinely 
        disadvantage certain social groups?
        \item \textbf{Fairness Repair:} If unfair, can we fix it (e.g. 
        with more optimization goals)?
    \end{itemize}
    \item \textbf{Knowledge:} How can we best 
    incorporate other knowledge sources (human, ensemble)?
    \item \textbf{Human Acceptance:} Will humans 
    accept recommendations, and do explanations change their 
    minds?
    \item \textbf{Human vs. Machine:} How do human vs. 
    machine explanations for "good" configurations compare?
\end{itemize}
\vspace{-0.2cm}
\noindent\rule{\linewidth}{0.4pt}
\end{tabular} 
\end{minipage}
\hfill
\begin{minipage}[t]{0.48\textwidth}
\begin{tabular}{@{} p{\columnwidth} @{}}
\hline 

{\bf C. Industrial Deployment \& Adoption} \rev{\textsc{[n]}}: 
Bridging the gap-research and practice. \\
 
\addlinespace[1pt]
\begin{itemize}[leftmargin=*, noitemsep, topsep=1pt, parsep=0pt]
    \item \textbf{Problem Identification:} How can we identify and 
    validate high-impact industrial problems where MSR 
    optimization techniques would provide the most value?
    \item \textbf{Case Studies:} What are the main barriers to 
    applying these methods to real-world industrial systems, and 
    how can they be overcome?
    \item \textbf{Data Collection:} How can we incentivize and manage 
    the collection of new industrial (perhaps proprietary) 
    optimization datasets for MOOT?
    \item \textbf{Education:} How to  introduce 
    newbies (in industry and academia) to these methods?
    \item \textbf{Deployment:} What tools \& tutorials are needed for widespread use of all this ?
\end{itemize}
\vspace{-0.2cm}
\noindent\rule{\linewidth}{0.4pt}
~\\

{\bf D. LLMs \& Emerging AI Technologies} \rev{\textsc{[e]}}: 
Foundation models and optimization. \\

\addlinespace[1pt]
\begin{itemize}[leftmargin=*, noitemsep, topsep=1pt, parsep=0pt]
\item \textbf{Distillation:} Figure~\ref{results} suggests that a few
labels can replace a large set. Could this be used to guide LLM distillation?  (finding small useful models
inside larger models?) 
    \item \textbf{Configuration Generation:} Can LLMs generate 
    effective initial configurations or suggest promising search 
    directions based on natural language descriptions?
    \item \textbf{Explanation Translation:} Can LLMs translate 
    technical optimization results into domain-specific 
    recommendations practitioners understand?
    \item \textbf{Prompt Engineering as Configuration:} Can MOOT 
    techniques optimize LLM prompts as multi-objective 
    configuration problems?
    \item \textbf{LLM-Generated Surrogates:} Can LLMs learn to 
    approximate expensive evaluation functions and serve as fast 
    surrogates during optimization?
    \item \textbf{Natural Language Constraints:} Can practitioners 
    specify configuration constraints in natural language, with 
    LLMs translating to formal specifications?
    \item \textbf{Meta-Learning:} Can LLMs identify which 
    optimization strategies work best for new problems based on 
    problem descriptions?
\end{itemize}
\vspace{-0.2cm}
\noindent\rule{\linewidth}{0.4pt}
\\

{\bf E. Solution Quality \& Generalizability} \rev{\textsc{[n]}}: 
Understand solution reliability.\\
  
\begin{itemize}[leftmargin=*, noitemsep, topsep=1pt, parsep=0pt]
    \item \textbf{Generalization:} Any commonalities in MOOT-generates solutions?
    \item \textbf{Transferability:} Can we transfer 
    knowledge between tasks, or must we start fresh?
    \item \textbf{Stability:} How stable are 
    conclusions across stochastic multi-objective algorithms?
    \item \textbf{Robustness:}   Robustness
     of results to 
    noise, incomplete data, changing conditions?
    \item \textbf{Uncertainty:} Can we learn  confidence intervals around our conclusions?
    \item \textbf{Failure Prediction:} Can data 
    features predict optimization failure in advance?
    \item \textbf{Constraints:} How to automatically detect and handle infeasible configurations?
\end{itemize}
\vspace{-0.2cm}
\noindent\rule{\linewidth}{0.4pt}
~\\

{\bf F. Future Directions} \rev{\textsc{[e]}}: 
Exploring new domains and advanced concepts. \\
  
\begin{itemize}[leftmargin=*, noitemsep, topsep=1pt, parsep=0pt]
    \item \textbf{Temporal:} 
    Reasoning across time. How to leverage the past? How (and when) to unlearn (to forget some of the past)?
    \item \textbf{Philosophical:} Is   solution space 
    "flat" (many "near-maxima") so no single "truth"? 
      \item \textbf{Artificial General Intelligence:} If we can learn,
    optimize, trade off goals, transfer knowledge to new  domains-- when does this become  AGI?
\end{itemize}
\vspace{-0.2cm}
\noindent\rule{\linewidth}{0.4pt}
\end{tabular} 
\end{minipage}
\end{table*}

\section{Discussion}



MOOT is open: new data sets are welcome via pull request.
We   plan (a) annual ICSE research events  on empirical optimization in SE, where results must generalize across MOOT data; and
(b)~tutorials at major SE venues (ICSE, FSE, ASE, MSR) on using MOOT-style tools.
We have also arranged for 
expedited publication of MOOT work in the \emph{Automated Software Engineering} journal (either as full papers or three-page tools/tutorials/registered reports\footnote{See \url{https://ause-journal.github.io/cfp.html}}
).
To qualify, mention ``MOOT'' in the title or abstract and use MOOT data.
\rev{We recommend: randomize row order before experiments; report results over multiple repeats; use nonparametric statistics for comparisons.}

Depending on how we use it, 
 MOOT 
could be a 
{\em catalyst}
or {\em cage}
for future research.
Used thoughtfully, it could broaden the questions we ask and the answers we
trust. But if used carelessly (obsessively, mindlessly) then
MOOT could suffer the same lifecycle as many other  repos: 
\begin{enumerate}
    \item \textit{Rejected} — “Find Data? That will never happen.”
    \item \textit{Respected} — “Fine\dots it helps sometimes.”
    \item \textit{Expected} — “You must compare to this baseline.”
    \item \textit{Exhausted} — the research graveyard where innovation stalls; 
          results become repetitive and derivative.
\end{enumerate}
To avoid the trap of the research graveyard,  
we should discourage papers that only offer minor percentage gains over existing baselines.
Instead, we should encourage work that, for example:
\begin{itemize}
    \item introduces {new tasks or domains}; or
    \item {challenges or expands} current assumptions; or
    \item {extends MOOT itself} (data, tooling, research directions, scripts, organization, or community).
\end{itemize}

\bibliography{refs}
\bibliographystyle{abbrvnat}
\end{document}